\documentclass[aps,prl,twocolumn,groupedaddress,amsmath]{revtex4-1}

\usepackage{color}
\usepackage{graphicx}
\usepackage{dcolumn}
\usepackage{bm}

\begin{document}


\title{Dynamics of a thin liquid film with surface rigidity and spontaneous curvature}


\author{Michael H. K\"opf}\email{m.koepf@uni-muenster.de}
\author{Svetlana V. Gurevich}
\author{Thomas Wulf}
\altaffiliation{Motion Science, University of M\"unster
Horstmarer Landweg 62b, D-48149 M\"unster}
\author{Rudolf Friedrich}
\affiliation{%
Institute for Theoretical Physics, University of M\"unster, Wilhelm-Klemm-Str.~9, D-48149 M\"unster, Germany}

\date{\today}

\begin{abstract}
The effect of rigid surfaces on the dynamics of thin liquid films which are amenable to
the lubrication approximation is considered. It is shown that the Helfrich 
energy of the layer gives rise to additional terms in the time-evolution equations 
of the liquid film. 
The dynamics is found to depend on the absolute value of the 
spontaneous curvature, irrespective of its sign. 
Due to the additional terms, a 
novel finite wavelength instability of flat rigid interfaces can be observed. 
Furthermore, the dependence of the shape of a droplet on the bending rigidity as 
well as on the spontaneous curvature is discussed. 
\end{abstract}

\pacs{68.15.+e, 47.20.Dr, 87.16.D-, 68.03.Cd}%

\maketitle

\section{Introduction}
Surfactant mono- and multilayers at a free interface, which are most commonly
formed by lipid molecules, exhibit extremely rich phase diagrams with an
abundance of different ordering states
\cite{KMD_RevModPhys_99,Moe_AnnuRevPhysChem_90}. Besides their fascinating
thermodynamics, they have been intensively studied as model systems for
biological
membranes which are composed of phospholipid bilayers \cite{NTN_Biochimica_00}.
Furthermore, floating surfactant layers are important for technical applications
as they can be
transferred onto solid substrates utilizing, e.g., Langmuir-Blodgett transfer, allowing
for a controlled coating with an arbitrary number of molecular layers and even
for the self-organized nano-patterning of surfaces
\cite{GCF_Nature_00,HJ_NatMater_05,KGFC_Langmuir_10,KGF_PRE_10}.

Depending on the chemical and thermodynamical properties of the used material, 
surfactant mono- and multilayers are rigid to some extent. The consequences are
twofold: On the one hand, the layer has a certain bending rigidity, so that 
additional work has to be surmounted in order to deform the surface. This
bending rigidity is measurable by x-ray scattering and depends strongly on the
surfactant material and the thermodynamic phase of the layer
\cite{DBB_PNAS_05,MDLS_EPL_04}. On the 
other hand, the layer can exhibit \emph{spontaneous curvature} due to
different
chemical properties of the head- and tailgroups of the surfactant molecules \cite{Safran}. 

One can expect these effects to be even more pronounced if the liquid support of such a surfactant layer becomes a very thin film, as is the case for example in coating processes, so that interfacial influences become increasingly important.
The dynamics of thin liquid films is accurately described within the framework
of the lubrication approximation \cite{ODB_RevModPhys_97,CM_RevModPhys_09}, which
has been used extensively throughout the literature to describe films covered
with layers of soluble- and insoluble surfactants
\cite{KGF_EPL_09,CM_PhysFluids_06,WGC_PhysFluids_94}. However, the focus of
these investigations has been on the effect of surfactant induced surface
tension gradients or on monolayer thermodynamics, whereas surface rigidity is
commonly neglected. It is the purpose of the present paper to generalize the
lubrication approximation of thin liquid films in this respect by incorporation
of the 
bending rigidity and the spontaneous curvature of a surface layer. 
Following this approach, we shall find that the surface rigidity gives rise to 
a pair of antagonistic effects: While short wavelength fluctuations are suppressed by the bending rigidity, the spontaneous curvature can lead to instabilities at intermediate wavelengths.
\section{Thin liquid films with rigid surface layers}
In the spirit of Helfrich, the energy of a rigid interfacial layer can be written as the sum of integrals over the mean curvature and the Gaussian curvature of the surface \cite{He_ZNaturforsch_73,Safran}. According to the Gauss-Bonnet theorem, the latter integral is a topological quantity, dependent only on the genus $p$ of the considered surface $S$. Denoting the principal curvatures of $S$ by $\kappa_1$ and $\kappa_2$, the spontaneous curvature by $\kappa_0$, and the surface tension by $\sigma$, the energy $E_\mathrm{r}$ due to the rigidity of a surface $S$ is accordingly obtained as
\begin{equation}
E_\mathrm{r} = \frac{k_\mathrm{c}}{2}\int\!\mathrm{d}S\, \left(\kappa_1+\kappa_2-\kappa_0\right)^2 + 2\pi k_\mathrm{g} (1-p).\label{eq:freeenergy}
\end{equation}
The positive material constants $k_\mathrm{c}$ and $k_\mathrm{g}$ are the bending 
rigidity and the modulus of the Gaussian curvature, respectively. 
\begin{figure}
\centering
\includegraphics[width=.45\textwidth]{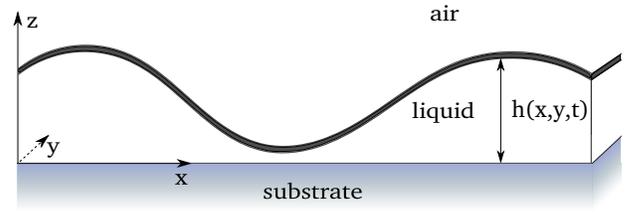}
\caption{\label{fig:drawing}Schematic drawing of a thin liquid film with a rigid surface layer. The shape of the film is described by its height profile $h(\bm{x},t)$, which gives the film thickness above point $\bm{x}$ at time $t$.}
\end{figure}
Here, we will focus on thin films which are entirely covered by a homogeneous surfactant layer without defects, that is, the case of constant surface topology. 
If the liquid film is described by its height profile $h(\bm{x},t)$ as shown in figure \ref{fig:drawing}, the Helfrich energy can be written a a functional $E_\mathrm{r}=E_\mathrm{r}[h]$.

In equilibrium there has to be a balance of forces $\delta W=0$ at the surface, where
\begin{equation}
\delta W = \int\!\mathrm{d} S \delta h \left\{p -\sigma(\kappa_1+\kappa_2) - 
\pi_\mathrm{d} + \frac{\delta E_\mathrm{r}}{\delta h}\right\}.
\label{eq:forcebalance}
\end{equation}
Here $p$ is the pressure in the liquid phase and $\pi_\mathrm{d}$ denotes the disjoining pressure describing the interaction between substrate and liquid which has to be taken into account for very thin films. 
To calculate the functional derivative of $E_\mathrm{r}$, we use the 
fact, that one can rewrite the mean curvature as
$\kappa_1+\kappa_2=-\nabla \cdot \bm{\hat{n}}$, where 
\begin{equation*}
\bm{\hat{n}} = \frac{1}{a^{1/2}}\left(\begin{array}{c}-\partial_x h \\ -\partial_y h \\ 1 \end{array}\right) \quad \textrm{with} \quad a=1+\left(\nabla h\right)^2
\end{equation*}
is the unit normal of the water surface, pointing into the vapor phase. We find
\begin{align}
\frac{\delta E_\mathrm{r}}{\delta h}=\,& k_\mathrm{c}\nabla \cdot \bigg\{\left(\nabla\cdot\bm{\hat{n}}+\kappa_0\right)\bigg[ a^{1/2}\left(\nabla\cdot\frac{\bm{\hat{n}}}{a}\right)\left(\nabla h\right) \notag\\ 
& + \frac{2\left(\Delta h\right)\bm{\hat{n}}-\left(\nabla\cdot\bm{\hat{n}}+\kappa_0\right)\left(\nabla h\right)}{2a^{1/2}}-\frac{\left(\nabla h\right)\left(\Delta h\right)}{a}  \bigg] \notag\\
& -\nabla\left(a^{-1}\left(\nabla\cdot\bm{\hat{n}}+\kappa_0\right)\right)\bigg\}. \label{eq:funcderiv}
\end{align}
Introducing the scales $h_0$, $l_0$, and $t_0$ for height, length, and time, one
can derive the time evolution equation for the nondimensionalized height profile $H(\bm{X},T):=h(\bm{x}/l_0,t/t_0)/h_0$ within the lubrication approximation. To this end, we follow precisely the steps in chapter II.B of the derivation by Oron et al.~\cite{ODB_RevModPhys_97}, with the difference, that our force balance equation contains the additional contribution $\delta E_\mathrm{r}/\delta h$. Expanding the functional derivative \eqref{eq:funcderiv} in powers 
of the parameter $\epsilon=h_0/l_0$, which is small for thin liquid films, one obtains
\begin{align*}
\frac{\delta E_\mathrm{r}}{\delta h} &= \frac{k_\mathrm{c}}{2}\nabla\cdot\bigg\{ \frac{2\epsilon}{l_0^3}\nabla\Delta H - \frac{\epsilon \kappa_0^2}{l_0}\nabla H  \\
&\phantom{=}\, -\frac{4 \epsilon^2\kappa_0}{l_0^2}\left[(\Delta H)\nabla H - \frac{1}{2}\nabla(\nabla H)^2\right]  \\
&\phantom{=}\, +\frac{5\epsilon^3}{l_0^3}(\Delta H)^2\nabla H \\ 
&\phantom{=}\, - \frac{\epsilon^3}{l_0^3}\nabla\left[3(\Delta H)(\nabla H)^2 + 2(\nabla H)\cdot\nabla (\nabla H)^2\right] \bigg\}.
\end{align*}
Inserting this result into the force balance equation
\eqref{eq:forcebalance} and expanding the remaining terms in powers of $\epsilon$ as well, we obtain, after dropping terms of higher orders in $\epsilon$, the following formula, which is a generalization of equation (2.24b) in \cite{ODB_RevModPhys_97}:
\begin{align*}
 -P + \Pi = \bar{\mathrm{C}}_\mathrm{I}^{-1} \Delta H + \bar{\mathrm{C}}_\mathrm{II}^{-1}\Delta^2 H,
\end{align*}
where $P=\epsilon h_0 p / (\mu U_0)$ and $\Pi=\epsilon h_0 \pi_\mathrm{d} / 
(\mu U_0)$  with characteristic velocity $U_0=l_0/t_0$ are the 
nondimensionalized pressure and disjoining pressure and 
\begin{equation}
\bar{\mathrm{C}}_\mathrm{I}^{-1} = \frac{\epsilon^3 }{U_0 \mu}\left(\sigma-\frac{k_\mathrm{c}\kappa_0^2}{2}\right) \quad \textrm{and} \quad \bar{\mathrm{C}}_\mathrm{II}^{-1}=\frac{\epsilon^3}{U_0 \mu} \frac{k_\mathrm{c}}{2 l_0^2}. \label{eq:cadefs}
\end{equation}
With this result at hand, one can follow the remaining steps of
the standard derivation in \cite{ODB_RevModPhys_97} to obtain the 
time evolution equation
\begin{equation}
\partial_T H = -\nabla\cdot\left\{\frac{H^3}{3}\nabla\left[\bar{\mathrm{C}}_\mathrm{I}^{-1}
\Delta H - \bar{\mathrm{C}}_\mathrm{II}^{-1}\Delta^2 H -\Pi(H)\right] \right\}. \label{eq:Hevol}
\end{equation}
It is a noteworthy result, that the thin film dynamics does not depend on the 
sign of the spontaneous curvature, but only on its absolute value.

Typical values for the bending modulus and the spontaneous curvature are $k_\mathrm{c}\approx 10^{-19}\,\mathrm{J}$ \cite{DBB_PNAS_05,MDLS_EPL_04} 
and $|\kappa_0|\approx 10^7\,\mathrm{m}^{-1}$ - $10^9\,\mathrm{m}^{-1}$ \cite{KCF_Biochemistry_05}, respectively. One can
thus roughly estimate $k_\mathrm{c}\kappa_0^2/2\approx 10^{-5}\,\mathrm{J}/\mathrm{m}^2$ - $10^{-1}\,\mathrm{J}/\mathrm{m}^2$. Comparing
this range of values to the surface tension of water in absence of any surfactant, $\sigma_\mathrm{abs}\approx 72\cdot 10^{-2}\,\mathrm{J}/\mathrm{m}^2$, we note
that the contribution due to spontaneous curvature can be of the same order 
of magnitude or even larger. The latter implies that spontaneous curvature 
can lead to a \emph{negative effective surface tension}. Assuming $l_0\approx 10^{-7}\,\mathrm{m}$ - $10^{-6}\,\mathrm{m}$ one can further estimate $k_\mathrm{c}/(2l_0^2)\approx 10^{-7}\,\mathrm{J}/\mathrm{m}^{2}$ - $10^{-5}\,\mathrm{J}/\mathrm{m}^{2}$. Thus, $\bar{\mathrm{C}}_\mathrm{II}^{-1}$
will in most cases be orders of magnitude smaller than $\bar{\mathrm{C}}_\mathrm{I}^{-1}$.
\section{Linear stability of flat films}
Any homogeneous film profile $H(\bm{X},T)=\hat{H}=\mathrm{const.}$ solves 
equation \eqref{eq:Hevol}. Considering small perturbations $\eta(\bm{X},T)$ of 
a flat film, we insert $H(\bm{X},T)=\hat{H}+\eta(\bm{X},T)$ into equation
\eqref{eq:Hevol} and keep only terms linear in $\eta$. Using the plane 
wave ansatz $\eta\sim\exp(\mathrm{i}\bm{k}\cdot\bm{X}+\lambda T)$ the dispersion
relation $\lambda(k)$ is obtained as
\begin{equation}
\lambda(k) = -\frac{\hat{H}^3}{3}k^2\left[\hat{\Pi}' + \bar{\mathrm{C}}_\mathrm{I}^{-1}k^2 + \bar{\mathrm{C}}_\mathrm{II}^{-1}k^4 
\right], \label{eq:dispersion}
\end{equation}
where we use the shorthand notation $\hat{\Pi}':=\partial_H \Pi(H)\big|_{H=\hat{H}}$ and $k:=|\bm{k}|$.
\begin{figure}
\centering
\includegraphics[width=.48\textwidth]{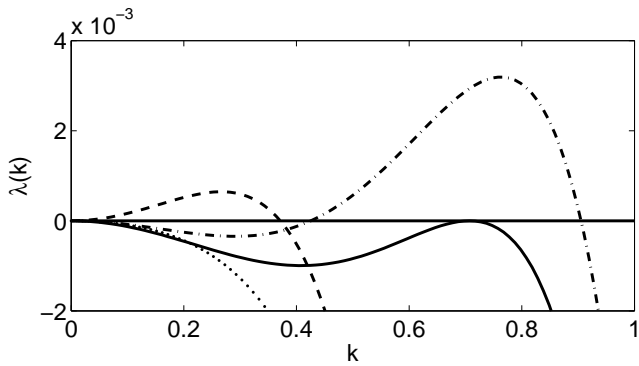}
\caption{\label{fig:lsa}Dispersion relation $\lambda(k)$ (see equation \eqref{eq:dispersion}) for four different parameter sets:
$\bar{\mathrm{C}}_\mathrm{I}^{-1}=-0.1$, 
$\bar{\mathrm{C}}_\mathrm{II}^{-1}=0.1$, $\hat{\Pi}'=0.025$ (solid line),
$\bar{\mathrm{C}}_\mathrm{I}^{-1}=-0.1$, $\bar{\mathrm{C}}_\mathrm{II}^{-1}=0.1$, $\hat{\Pi}'=0.0148$ (dash-dotted line), $\bar{\mathrm{C}}_\mathrm{I}^{-1}=0.1$, $\bar{\mathrm{C}}_\mathrm{II}^{-1}=0.1$, $\hat{\Pi}'=-0.0157$ (dashed line),
and $\bar{\mathrm{C}}_\mathrm{I}^{-1}=0.1$,
$\bar{\mathrm{C}}_\mathrm{II}^{-1}=0.1$, $\hat{\Pi}'=0.0148$ (dotted line).}
\end{figure}
This result complies with the physical intuition, that, for vanishing 
spontaneous curvature, the bending rigidity will damp short wavelength
fluctuations and thus has a stabilizing influence. However, if
$\bar{\mathrm{C}}_\mathrm{I}^{-1}$ takes negative values, this can lead to an
instability for intermediate wavenumbers: For 
$\bar{\mathrm{C}}_\mathrm{I}^{-1} < 0$ and $\hat{
 \Pi}'=\bar{\mathrm{C}}_\mathrm{I}^{-2}/(4\bar{\mathrm{C}}_\mathrm{II}^{-1})$ 
the eigenvalue $\lambda(k_\mathrm{cr})$ of the mode $k_\mathrm{cr}=\sqrt{-\bar{\mathrm{C}}_\mathrm{I}^{-1}/
(2\bar{\mathrm{C}}_\mathrm{II}^{-1}})$ crosses the imaginary axis (solid line in figure \ref{fig:lsa}), so that for  
$0<\hat{ \Pi}'<\mathrm{C}_\mathrm{I}^{-2}/(4\bar{\mathrm{C}}_\mathrm{II}^{-1})$
and $\bar{\mathrm{C}}_\mathrm{I}^{-1}<0$ there is a band of unstable modes
$k_1<k<k_2$ (dash-dotted line in figure \ref{fig:lsa}) with
\begin{equation*}
k_{1,2}=\sqrt{\frac{-\bar{\mathrm{C}}_\mathrm{I}^{-1}\mp\sqrt{\bar{\mathrm{C}}_\mathrm{I}^{-2} - 4 \bar{\mathrm{C}}_\mathrm{II}^{-1}\hat{ \Pi}'}}{2\bar{\mathrm{C}}_\mathrm{II}^{-1}}}.
\end{equation*}
Obviously $k_1=0$ for negative $\hat{\Pi}'$ and arbitrary values of $\bar{\mathrm{C}}_\mathrm{I}^{-1}$ (dashed line in figure \ref{fig:lsa}).
The wavenumber $k_\mathrm{max}$ corresponding to the maximum of $\lambda(k)$, 
that is, the most unstable mode, is given by
\begin{equation*}
k_\mathrm{max}=\sqrt{\frac{-\bar{\mathrm{C}}_\mathrm{I}^{-1}+\sqrt{\bar{\mathrm{C}}_\mathrm{I}^{-2}-3\bar{\mathrm{C}}_\mathrm{II}^{-1}\hat{\Pi}'}}{3\bar{\mathrm{C}}_\mathrm{II}^{-1}}}
\end{equation*}
with
\begin{align*}
\lambda(k_\mathrm{max})=& -\frac{\hat{H}^3\left(-\bar{\mathrm{C}}_\mathrm{I}^{-1}+\sqrt{\bar{\mathrm{C}}_\mathrm{I}^{-2}-3\bar{\mathrm{C}}_\mathrm{II}^{-1}\hat{\Pi}'}\right)}{81 \bar{\mathrm{C}}_\mathrm{II}^{-2}} \\
& \times
\left( 6 \bar{\mathrm{C}}_\mathrm{II}^{-1}\hat{\Pi}' -\bar{\mathrm{C}}_\mathrm{I}^{-2} +  \bar{\mathrm{C}}_\mathrm{I}^{-1}\sqrt{\bar{\mathrm{C}}_\mathrm{I}^{-2}-3\bar{\mathrm{C}}_\mathrm{II}^{-1}\hat{\Pi}'}\right).
\end{align*}
For $\hat{\Pi}>0$, $\mathrm{C}_\mathrm{I}^{-1}>0$, and
$\mathrm{C}_\mathrm{II}^{-1}>0$, the whole spectrum is damped (dotted line in 
figure \ref{fig:lsa}).

\begin{figure}
\centering
\includegraphics[width=.49\textwidth]{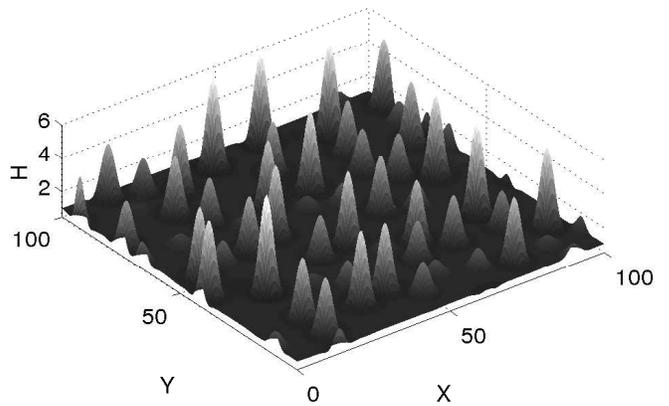}
\caption{\label{fig:2Ddrops}Snapshot of a numerical simulation of the breakup of 
a homogeneous film with $\hat{H}=1.2$, $\mathrm{C}^{-1}_\mathrm{I}=-0.1$, $\mathrm{C}^{-1}_\mathrm{II}=0.1$.}
\end{figure}
A numerical simulation \footnote{Both, one- and two-dimensional direct numerical
integration was carried out using an embedded Runge-Kutta scheme of order 4 (5)
for time-stepping \cite{NRC} and finite differences of second order for 
spatial discretization. The gridsize was $256$ and $192\times 192$ in the 1D 
and 2D cases, respectively.}
of a randomly perturbed flat film of height $\hat{H}=1.2$ with
$\bar{\mathrm{C}}_\mathrm{I}^{-1}=-0.1$ and 
$\bar{\mathrm{C}}_\mathrm{II}^{-1}=0.1$ was carried out
using the disjoining pressure
$
\Pi(H)=AH^{-3}\left(1-H^{-3} \right) 
$
with Hamaker constant $A=6.5\cdot 10^{-2}$. This is precisely the situation 
corresponding to the dash-dotted line in figure \ref{fig:lsa}. 
The flat film quickly breaks up into small droplets which 
then undergo a process of coarsening, as is commonly observed 
after film rupture. 

Remarkably, however, these results predict, that rigid layers with
spontaneous curvature can be used to destabilize very thin films, 
which would remain flat in absence of the surfactant layer, and
to obtain droplets of extremely small volumes after film breakup.
\section{The shape of a drop covered with a rigid layer}
\begin{figure}
\centering
\includegraphics[width=.49\textwidth]{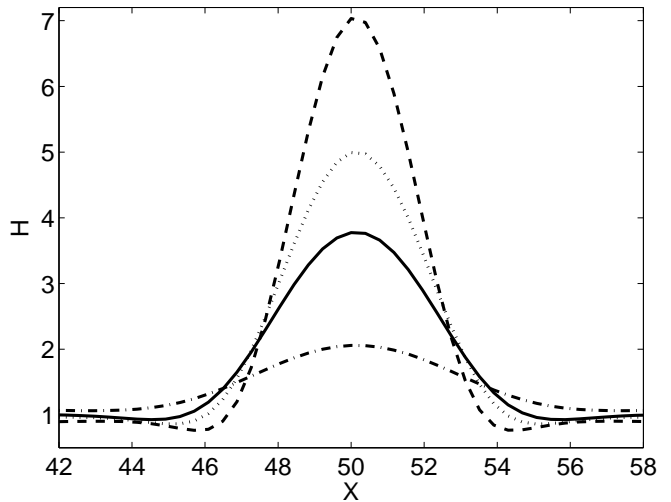}
\caption{\label{fig:shapeslv}The different shapes of a droplet obtained for $\bar{\mathrm{C}}^{-1}_\mathrm{II}=0.1$ and  $\bar{\mathrm{C}}^{-1}_\mathrm{I}=0$ (dash-dotted line), $\bar{\mathrm{C}}^{-1}_\mathrm{I}=-0.05$ (solid line), $\bar{\mathrm{C}}^{-1}_\mathrm{I}=-0.07$ (dotted line), $\bar{\mathrm{C}}^{-1}_\mathrm{I}=-0.1$ (dashed line). The overall volume of liquid was equal in all four cases. The image shows a closeup of a steady state on a periodic domain of size $L=100$.}
\end{figure}
It was shown in the preceding section, that surface rigidity 
affects the linear stability of flat films. Now, we want to 
investigate how the rigidity affects the shape of the droplets that
are formed during the rupture of an unstable film. To this end, we 
have considered steady state solutions of equation \eqref{eq:Hevol} on a one-dimensional domain which were obtained by relaxation, that is, 
direct numerical integration of equation \eqref{eq:Hevol}. When a steady state 
was obtained for one set of parameters, it has been used as initial condition for the subsequent run. Thus, the overall volume of liquid within the integration domain was constant throughout a whole 
parameter scan.

First, $\bar{\mathrm{C}}_\mathrm{II}^{-1}=0.1$ was held constant
while $\bar{\mathrm{C}}_\mathrm{I}^{-1}$ was varied from $0$ to $-0.1$. The obtained
shape of the droplet depends sensitively on $\bar{\mathrm{C}}_\mathrm{I}^{-1}$,
as is shown in figure \ref{fig:shapeslv}. The drop becomes steeper and develops 
more and more pronounced undershoots below the precursor height $H=1$ as the parameter is decreased.

\begin{figure}
\centering
\includegraphics[width=.49\textwidth]{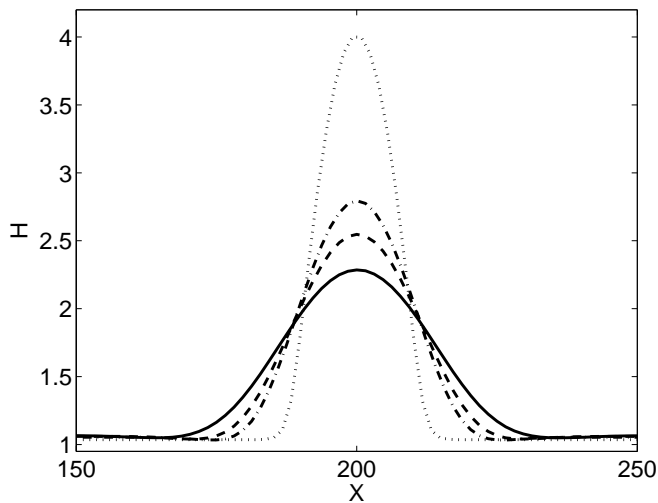}
\caption{\label{fig:shapes2}The different shapes of a droplet obtained for $\bar{\mathrm{C}}^{-1}_\mathrm{I}=0.1$ and  $\bar{\mathrm{C}}^{-1}_\mathrm{II}=0$ (dotted line), $\bar{\mathrm{C}}^{-1}_\mathrm{II}=10$ (dash-dotted line), $\bar{\mathrm{C}}^{-1}_\mathrm{II}=30$ (dashed line), $\bar{\mathrm{C}}^{-1}_\mathrm{II}=50$ (solid line). The overall volume of liquid was equal in all four cases. The image shows a closeup of a steady state on a periodic domain of size $L=100$.}
\end{figure}
In a second parameter scan $\bar{\mathrm{C}}_\mathrm{I}^{-1}=0.1$ was fixed 
while $\bar{\mathrm{C}}_\mathrm{II}^{-1}$ was varied from $0$ to $50$, as is 
shown in figure \ref{fig:shapes2}. The results indicate that the scaled bending
modulus $k_\mathrm{c}/(2l_0^2)$ needs to be $10$ - $100$ times higher than the effective surface tension in order to change the shape of the droplet
significantly.
\section{Conclusions \& Outlook}
A generalization of the equations of thin film flow to the case
of rigid surface layers with spontaneous curvature has been presented. 
Our results show, that the spontaneous curvature can lead to a significantly lower 
and even negative effective surface tension, yielding a new finite wavelength 
instability resulting in the breakup of very thin flat films. It turned out, 
that only
the absolute value and not the sign of the spontaneous curvature is important
in the lubrication regime.

We have shown that the sensitivity of steady droplet solutions on $\bar{\mathrm{C}}_\mathrm{II}^{-1}$ 
is very low. Combined with the fact that it is very small in most physical situations, 
we conclude that its most important contribution is the stabilization of short 
wavelength fluctuations in the case of negative  $\bar{\mathrm{C}}_\mathrm{I}^{-1}$.

It would be of interest to further extend the model to the case of surfactant 
layers with evolving topology, that is, to layers that can break up and form 
holes, leading to contributions $\sim k_\mathrm{g}$ to the surface energy. This 
could also be applicable to binary surface layers consisting of material in two 
different thermodynamical phases, if only one phase has significant rigidity.
\begin{acknowledgments}
This work was supported by the Deutsche Forschungsgemeinschaft within SRF TRR 61. \end{acknowledgments}

\begin{thebibliography}{10}%
\makeatletter
\providecommand \@ifxundefined [1]{%
 \ifx #1\undefined \expandafter \@firstoftwo
 \else \expandafter \@secondoftwo
\fi
}%
\providecommand \@ifnum [1]{%
 \ifnum #1\expandafter \@firstoftwo
 \else \expandafter \@secondoftwo
\fi
}%
\providecommand \enquote [1]{``#1''}%
\providecommand \bibnamefont  [1]{#1}%
\providecommand \bibfnamefont [1]{#1}%
\providecommand \citenamefont [1]{#1}%
\providecommand\href[0]{\@sanitize\@href}%
\providecommand\@href[1]{\endgroup\@@startlink{#1}\endgroup\@@href}%
\providecommand\@@href[1]{#1\@@endlink}%
\providecommand \@sanitize [0]{\begingroup\catcode`\&12\catcode`\#12\relax}%
\@ifxundefined \pdfoutput {\@firstoftwo}{%
 \@ifnum{\z@=\pdfoutput}{\@firstoftwo}{\@secondoftwo}%
}{%
 \providecommand\@@startlink[1]{\leavevmode}%
 \providecommand\@@endlink[0]{}%
}{%
 \providecommand\@@startlink[1]{%
  \leavevmode
  \pdfstartlink
   attr{/Border[0 0 1 ]/H/I/C[0 1 1]}%
   user{/Subtype/Link/A<</Type/Action/S/URI/URI(#1)>>}%
  \relax
 }%
 \providecommand\@@endlink[0]{\pdfendlink}%
}%
\providecommand \url  [0]{\begingroup\@sanitize \@url }%
\providecommand \@url [1]{\endgroup\@href {#1}{\urlprefix}}%
\providecommand \urlprefix [0]{URL }%
\providecommand \Eprint[0]{\href }%
\@ifxundefined \urlstyle {%
  \providecommand \doi [1]{doi:\discretionary{}{}{}#1}%
}{%
  \providecommand \doi [0]{doi:\discretionary{}{}{}\begingroup
  \urlstyle{rm}\Url }%
}%
\providecommand \doibase [0]{http://dx.doi.org/}%
\providecommand \Doi[1]{\href{\doibase#1}}%
\providecommand \bibAnnote [3]{%
  \BibitemShut{#1}%
  \begin{quotation}\noindent
    \textsc{Key:}\ #2\\\textsc{Annotation:}\ #3%
  \end{quotation}%
}%
\providecommand \bibAnnoteFile [2]{%
  \IfFileExists{#2}{\bibAnnote {#1} {#2} {\input{#2}}}{}%
}%
\providecommand \typeout [0]{\immediate \write \m@ne }%
\providecommand \selectlanguage [0]{\@gobble}%
\providecommand \bibinfo [0]{\@secondoftwo}%
\providecommand \bibfield [0]{\@secondoftwo}%
\providecommand \translation [1]{[#1]}%
\providecommand \BibitemOpen[0]{}%
\providecommand \bibitemStop [0]{}%
\providecommand \bibitemNoStop [0]{.\EOS\space}%
\providecommand \EOS [0]{\spacefactor3000\relax}%
\providecommand \BibitemShut [1]{\csname bibitem#1\endcsname}%
\bibitem{Moe_AnnuRevPhysChem_90}%
  \BibitemOpen
  \bibfield{author}{%
  \bibinfo {author} {\bibfnamefont{H.}~\bibnamefont{M\"ohwald}},\ }%
  \bibfield{booktitle}{%
  \emph{\bibinfo {booktitle} {Annual Review of Physical Chemistry}},\ }%
  \bibfield{journal}{%
  \bibinfo {journal} {Annu. Rev. Phys. Chem.}\ }%
  \textbf{\bibinfo {volume} {41}},\ \bibinfo {pages} {441} (\bibinfo {year} {1990})
  \bibAnnoteFile{NoStop}{Moe_AnnuRevPhysChem_90}%
\bibitem{KMD_RevModPhys_99}%
  \BibitemOpen
  \bibfield{author}{%
  \bibinfo {author} {\bibfnamefont{V.~M.}\ \bibnamefont{Kaganer}}, \bibinfo
  {author} {\bibfnamefont{H.}~\bibnamefont{M\"ohwald}},\ and\ \bibinfo {author}
  {\bibfnamefont{P.}~\bibnamefont{Dutta}},\ }%
  \bibfield{journal}{%
  \Doi{10.1103/RevModPhys.71.779}{\bibinfo {journal} {Rev. Mod. Phys.}}\ }%
  \textbf{\bibinfo {volume} {71}},\ \bibinfo {pages} {779} (\bibinfo {year}
  {1999})%
  \bibAnnoteFile{NoStop}{KMD_RevModPhys_99}%
\bibitem{NTN_Biochimica_00}%
  \BibitemOpen
  \bibfield{author}{%
  \bibinfo {author} {\bibfnamefont{J.~F.}\ \bibnamefont{Nagle}}\ and\ \bibinfo
  {author} {\bibfnamefont{S.}~\bibnamefont{Tristram-Nagle}},\ }%
  \bibfield{journal}{%
  \Doi{DOI: 10.1016/S0304-4157(00)00016-2}{\bibinfo {journal} {Biochim.~Biophys.~Acta}}\ }%
  \textbf{\bibinfo {volume} {1469}},\ \bibinfo {pages} {159 } (\bibinfo {year}
  {2000})%
  \bibAnnoteFile{NoStop}{NTN_Biochimica_00}%
\bibitem{GCF_Nature_00}%
  \BibitemOpen
  \bibfield{author}{%
  \bibinfo {author} {\bibfnamefont{M.}~\bibnamefont{{Gleiche}}}, \bibinfo
  {author} {\bibfnamefont{L.~F.}\ \bibnamefont{{Chi}}},\ and\ \bibinfo {author}
  {\bibfnamefont{H.}~\bibnamefont{{Fuchs}}},\ }%
  \bibfield{journal}{%
  \Doi{10.1038/35003149}{\bibinfo {journal} {Nature}}\ }%
  \textbf{\bibinfo {volume} {403}},\ \bibinfo {pages} {173} (\bibinfo {year}
  {2000})%
  \bibAnnoteFile{NoStop}{GCF_Nature_00}%
\bibitem{HJ_NatMater_05}%
  \BibitemOpen
  \bibfield{author}{%
  \bibinfo {author} {\bibfnamefont{J.}~\bibnamefont{Huang}}, \bibinfo {author}
  {\bibfnamefont{F.}~\bibnamefont{Kim}}, \bibinfo {author}
  {\bibfnamefont{A.~R.}\ \bibnamefont{Tao}}, \bibinfo {author}
  {\bibfnamefont{S.}~\bibnamefont{Connor}},\ and\ \bibinfo {author}
  {\bibfnamefont{P.}~\bibnamefont{Yang}},\ }%
  \bibfield{journal}{%
  \bibinfo {journal} {Nature Materials}\ }%
  \textbf{\bibinfo {volume} {4}},\ \bibinfo {pages} {896} (\bibinfo {year}
  {2005})%
  \bibAnnoteFile{NoStop}{HJ_NatMater_05}%
\bibitem{KGFC_Langmuir_10}%
  \BibitemOpen
  \bibfield{author}{%
  \bibinfo {author} {\bibfnamefont{M.~H.}\ \bibnamefont{K\"opf}}, \bibinfo
  {author} {\bibfnamefont{S.~V.}\ \bibnamefont{Gurevich}}, \bibinfo {author}
  {\bibfnamefont{R.}~\bibnamefont{Friedrich}},\ and\ \bibinfo {author}
  {\bibfnamefont{L.}~\bibnamefont{Chi}},\ }%
  \bibfield{journal}{%
  \bibinfo {journal} {Langmuir}\ }%
  \textbf{\bibinfo {volume} {26}},\ \bibinfo {pages} {10444} (\bibinfo {year}
{2010})%
  \bibAnnoteFile{NoStop}{KGFC_Langmuir_10}%
\bibitem{KGF_PRE_10}%
  \BibitemOpen
  \bibfield{author}{%
  \bibinfo {author} {\bibfnamefont{M.~H.}\ \bibnamefont{K\"opf}}, \bibinfo
  {author} {\bibfnamefont{S.~V.}\ \bibnamefont{Gurevich}},\ and\ \bibinfo
{author}
  {\bibfnamefont{R.}~\bibnamefont{Friedrich}},\ }%
  \bibfield{journal}{%
  \bibinfo {journal} {Phys. Rev. E}\ }%
  \textit{accepted} (\bibinfo {year}
{2010})%
  \bibAnnoteFile{NoStop}{KGF_PRE_10}%
\bibitem{MDLS_EPL_04}%
  \BibitemOpen
  \bibfield{author}{%
  \bibinfo {author} {\bibfnamefont{S.}~\bibnamefont{Mora}}, \bibinfo {author}
  {\bibfnamefont{J.}~\bibnamefont{Daillant}}, \bibinfo {author}
  {\bibfnamefont{D.}~\bibnamefont{Luzet}},\ and\ \bibinfo {author}
  {\bibfnamefont{B.}~\bibnamefont{Struth}},\ }%
  \bibfield{journal}{%
  \bibinfo {journal} {EPL (Europhysics Letters)}\ }%
  \textbf{\bibinfo {volume} {66}},\ \bibinfo {pages} {694} (\bibinfo {year}
  {2004})%
  \bibAnnoteFile{NoStop}{MDLS_EPL_04}%
\bibitem{DBB_PNAS_05}%
  \BibitemOpen
  \bibfield{author}{%
  \bibinfo {author} {\bibfnamefont{J.}~\bibnamefont{Daillant}}, \bibinfo
  {author} {\bibfnamefont{E.}~\bibnamefont{Bellet-Amalric}}, \bibinfo {author}
  {\bibfnamefont{A.}~\bibnamefont{Braslau}}, \bibinfo {author}
  {\bibfnamefont{T.}~\bibnamefont{Charitat}}, \bibinfo {author}
  {\bibfnamefont{G.}~\bibnamefont{Fragneto}}, \bibinfo {author}
  {\bibfnamefont{F.}~\bibnamefont{Graner}}, \bibinfo {author}
  {\bibfnamefont{S.}~\bibnamefont{Mora}}, \bibinfo {author}
  {\bibfnamefont{F.}~\bibnamefont{Rieutord}},\ and\ \bibinfo {author}
  {\bibfnamefont{B.}~\bibnamefont{Stidder}},\ }%
  \bibfield{journal}{%
  \bibinfo {journal} {Proceedings of the National Academy of Sciences of the
  United States of America}\ }%
  \textbf{\bibinfo {volume} {102}},\ \bibinfo {pages} {11639} (\bibinfo {year}
  {2005})%
  \bibAnnoteFile{NoStop}{DBB_PNAS_05}%
\bibitem{Safran}%
  \BibitemOpen
  \bibfield{author}{%
  \bibinfo {author} {\bibfnamefont{S.~A.}\ \bibnamefont{Safran}},\ }%
  \emph{\bibinfo {title} {Statisitical Thermodynamics of Surfaces, Interfaces,
  and Membranes}},\ Frontiers in Physics\ (\bibinfo {publisher}
  {Addison-Wesley},\ \bibinfo {address} {Reading, Massachusetts},\ \bibinfo
  {year} {1994})%
  \bibAnnoteFile{NoStop}{Safran}%
\bibitem{ODB_RevModPhys_97}%
  \BibitemOpen
  \bibfield{author}{%
  \bibinfo {author} {\bibfnamefont{A.}~\bibnamefont{Oron}}, \bibinfo {author}
  {\bibfnamefont{S.~H.}\ \bibnamefont{Davis}},\ and\ \bibinfo {author}
  {\bibfnamefont{S.~G.}\ \bibnamefont{Bankoff}},\ }%
  \bibfield{journal}{%
  \Doi{10.1103/RevModPhys.69.931}{\bibinfo {journal} {Rev. Mod. Phys.}}\ }%
  \textbf{\bibinfo {volume} {69}},\ \bibinfo {pages} {931} (\bibinfo {year}
  {1997})%
  \bibAnnoteFile{NoStop}{ODB_RevModPhys_97}%
\bibitem{CM_RevModPhys_09}%
  \BibitemOpen
  \bibfield{author}{%
  \bibinfo {author} {\bibfnamefont{R.~V.}\ \bibnamefont{Craster}}\ and\
  \bibinfo {author} {\bibfnamefont{O.~K.}\ \bibnamefont{Matar}},\ }%
  \bibfield{journal}{%
  \Doi{10.1103/RevModPhys.81.1131}{\bibinfo {journal} {Rev. of Mod. Phys.}}\ }%
  \textbf{\bibinfo {volume} {81}},\ \bibinfo {eid} {1131} (\bibinfo {year}
  {2009})
  \bibAnnoteFile{NoStop}{CM_RevModPhys_09}%
\bibitem{WGC_PhysFluids_94}%
  \BibitemOpen
  \bibfield{author}{%
  \bibinfo {author} {\bibfnamefont{A.}~\bibnamefont{{De Wit}}}, \bibinfo
  {author} {\bibfnamefont{D.}~\bibnamefont{Gallez}},\ and\ \bibinfo {author}
  {\bibfnamefont{C.~I.}\ \bibnamefont{Christov}},\ }%
  \bibfield{journal}{%
  \Doi{10.1063/1.868058}{\bibinfo {journal} {Phys. Fluids}}\ }%
  \textbf{\bibinfo {volume} {6}},\ \bibinfo {pages} {3256} (\bibinfo {year}
  {1994})%
  \bibAnnoteFile{NoStop}{WGC_PhysFluids_94}%
\bibitem{CM_PhysFluids_06}%
  \BibitemOpen
  \bibfield{author}{%
  \bibinfo {author} {\bibfnamefont{R.~V.}\ \bibnamefont{Craster}}\ and\
  \bibinfo {author} {\bibfnamefont{O.~K.}\ \bibnamefont{Matar}},\ }%
  \bibfield{journal}{%
  \Doi{10.1063/1.2180776}{\bibinfo {journal} {Phys. Fluids}}\ }%
  \textbf{\bibinfo {volume} {18}},\ \bibinfo {pages} {032103 (12 pages)}
  (\bibinfo {year} {2006})%
  \bibAnnoteFile{NoStop}{CM_PhysFluids_06}%
\bibitem{KGF_EPL_09}%
  \BibitemOpen
  \bibfield{author}{%
  \bibinfo {author} {\bibfnamefont{M.~H.}\ \bibnamefont{K\"opf}}, \bibinfo
  {author} {\bibfnamefont{S.~V.}\ \bibnamefont{Gurevich}},\ and\ \bibinfo
  {author} {\bibfnamefont{R.}~\bibnamefont{Friedrich}},\ }%
  \bibfield{journal}{%
  \bibinfo {journal} {EPL (Europhysics Letters)}\ }%
  \textbf{\bibinfo {volume} {86}},\ \bibinfo {pages} {66003}
  (\bibinfo {year} {2009})
  \bibAnnoteFile{NoStop}{KGF_EPL_09}%
\bibitem{He_ZNaturforsch_73}%
  \BibitemOpen
  \bibfield{author}{%
  \bibinfo {author} {\bibfnamefont{W.}~\bibnamefont{Helfrich}},\ }%
  \bibfield{journal}{%
  \bibinfo {journal} {Z. Naturforsch.}\ }%
  \textbf{\bibinfo {volume} {28 c}},\ \bibinfo {pages} {693} (\bibinfo {year}
  {1973})%
  \bibAnnoteFile{NoStop}{He_ZNaturforsch_73}%
\bibitem{KCF_Biochemistry_05}%
  \BibitemOpen
  \bibfield{author}{%
  \bibinfo {author} {\bibfnamefont{E.~E.}\ \bibnamefont{Kooijman}}, \bibinfo
  {author} {\bibfnamefont{V.}~\bibnamefont{Chupin}}, \bibinfo {author}
  {\bibfnamefont{N.~L.}\ \bibnamefont{Fuller}}, \bibinfo {author}
  {\bibfnamefont{M.~M.}\ \bibnamefont{Kozlov}}, \bibinfo {author}
  {\bibfnamefont{B.}~\bibnamefont{de~Kruijff}}, \bibinfo {author}
  {\bibfnamefont{K.~N.~J.}\ \bibnamefont{Burger}},\ and\ \bibinfo {author}
  {\bibfnamefont{P.~R.}\ \bibnamefont{Rand}},\ }%
  \bibfield{journal}{%
  \Doi{10.1021/bi0478502}{\bibinfo {journal} {Biochemistry}}\ }%
  \textbf{\bibinfo {volume} {44}},\ \bibinfo {pages} {2097} (\bibinfo {year}
  {2005})
  \bibAnnoteFile{NoStop}{KCF_Biochemistry_05}%
\bibitem{Note1}%
  \BibitemOpen
  \bibinfo {note} {Both, one- and two-dimensional direct numerical integration
  was carried out using an embedded Runge-Kutta scheme of order 4 (5) for
  time-stepping \cite {NRC} and finite differences of second order for spatial
  discretization. The gridsize was $256$ and $192\times 192$ in the 1D and 2D
  cases, respectively.}%
  \bibAnnoteFile{Stop}{Note1}%
\bibitem{NRC}%
  \BibitemOpen
  \bibfield{author}{%
  \bibinfo {author} {\bibfnamefont{W.~H.}\ \bibnamefont{Press}},\ }%
  \emph{\bibinfo {title} {Numerical Recipes in C}}\ (\bibinfo {publisher}
  {University Press},\ \bibinfo {address} {Cambridge},\ \bibinfo {year}
  {1999})%
  \bibAnnoteFile{NoStop}{NRC}%
\end{thebibliography}
\end{document}